\documentclass[twocolumn]{aastex61}
\usepackage{natbib}
\usepackage{amsmath,wasysym}

\revised{\today}
\submitjournal{ApJL}
\shorttitle{Magnetic fields in Hot Jupiters}
\shortauthors{Yadav and Thorngren}

\begin{document}

\title{Estimating the magnetic field strength in hot Jupiters}

\correspondingauthor{Rakesh K.Yadav}
\email{rakesh\_yadav@fas.harvard.edu}

\author[0000-0002-9569-2438]{Rakesh K. Yadav}
\affil{Department of Earth and Planetary Sciences, Harvard University, 20 Oxford St., Cambridge 02138, MA, USA}

\author{Daniel P. Thorngren}
\affiliation{Department of Physics, University of California, Santa
Cruz, USA}

\begin{abstract}

A large fraction of known Jupiter like exoplanets are inflated as compared to Jupiter. These ``hot" Jupiters orbit close to their parent star and are bombarded with intense starlight. Many theories have been proposed to explain their radius inflation and several suggest that a small fraction of the incident starlight is injected in to the planetary interior which helps to puff up the planet. How will such energy injection affect the planetary dynamo? In this Letter, we estimate the surface magnetic field strength of hot Jupiters using scaling arguments that relate energy available in planetary interiors to the dynamo generated magnetic fields. We find that if we take into account the energy injected in the planetary interior that is sufficient to inflate hot Jupiters to observed radii, then the resulting dynamo should be able  generate magnetic fields that are more than an order of magnitude stronger than the Jovian values. Our analysis highlights the potential fundamental role of the stellar light in setting the field strength in hot Jupiters.

\end{abstract}

\section{Introduction} \label{sec:intro}

The discovery and characterization of exoplanets have revolutionized planetary physics. Jupiter like exoplanets orbiting close to their parent star were among the first to be confirmed and they are generally referred to as ``Hot Jupiters" (HJs). Observations reveal that HJs have a surprisingly large range of radii and most are inflated as compared to Jupiter. This is particularly surprising since theory predicts that giant planets with a broad range of masses, going from $\approx0.3 M_J$ to $\approx13M_J$ (where $M_J$ is Jupiter mass), will quickly converge to Jupiter like radius ($R_J$) after few Gyrs \citep{burrows1997, marley2007}. Since most HJs are inflated, we may speculate that somehow HJs are either not cooling down as efficiently as we thought or they might have a constant supply of heat from some mechanism \citep{spiegel2013, komacek2017b}.

\begin{figure}[ht!]
\epsscale{1.18}\plotone{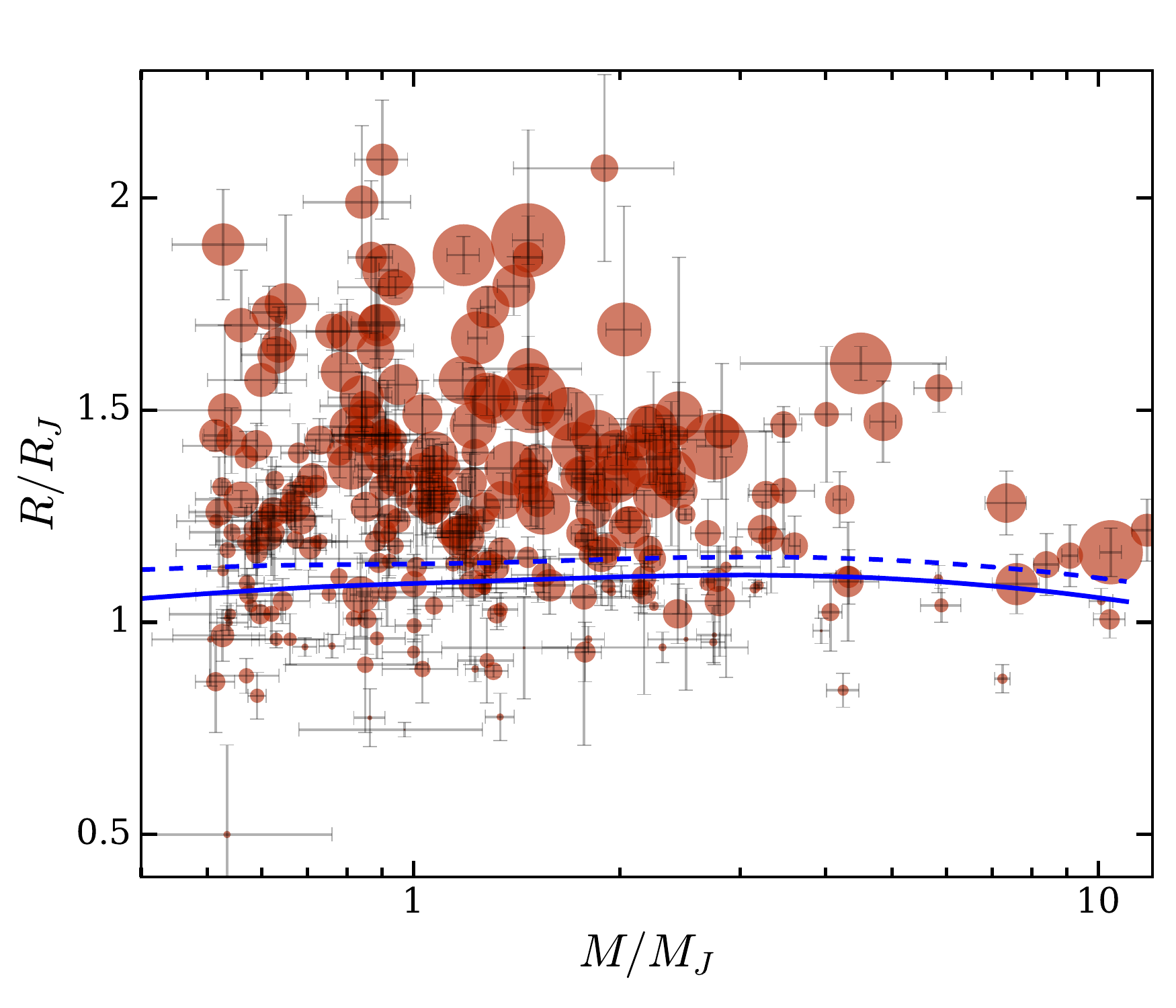}
\caption{Radius and mass (relative to Jupiter) of giant planets. The data is taken from exoplanet.eu \citep{schneider2011}, and the NASA exoplanet archive \citep{akeson2013}. We only chose exoplanets whose relative radius and mass uncertainties are below 50\%. The area of the data points scale as a function of the stellar heat flux at the planetary orbit. The smallest and largest values (in $erg\,s^{-1}\,cm^{-2}$) of incident stellar flux are about $10^{6}$ and $10^{10}$.  The blue solid and broken curves represent the radius predictions for solar-ratio H/He planets (no metals) of age 5 and 1 Gyrs and an incident stellar flux of $10^6$ \citep{thorngren2017}.\label{fig:R_M}}
\end{figure}

In Fig.~\ref{fig:R_M}, we plot the radius of giant exoplanets (restricted to 0.5$M_J$ to 12$M_J$) vs.~their mass, with the symbols sizes representing the incident stellar flux. It is evident that the degree of HJ inflation is roughly correlated with the amount of stellar heat flux available at the planetary orbits \citep{laughlin2011}. This idea is bolstered by recent work \citep{lopez2016, grunblatt2016, hartman2016} showing that HJs become more inflated as their parent stars become brighter with age. Therefore, regardless of the mechanism (for a recent summary, see \citealp{komacek2017b}) that channels stellar heat into the planetary interiors, it is safe to assume that HJs have a large pool of energy available in their interior that leads to radius inflation.

What happens to the planetary dynamo if the interiors of HJs are trapped in a highly energetic state for long times?  Multiple theories have been proposed in the past that predict the mean magnetic field in planets and stars using their physical properties (for a review, see \citealp{christensen2010}). Recently, \citet{christensen2009} found that a scaling law inferred from geodynamo simulations predicts the mean magnetic field strength in Earth, Jupiter, low-mass main-sequence stars, and T Tauri stars, to a good degree. This law relates the planetary luminosity with the planetary mean magnetic field. Motivated by these encouraging results, \citet{reiners2010} applied the scaling law to Jupiter like exoplanets with a mass range of $M_J$ to 13$M_J$ and predict a range of 10G to 100G, respectively, for the dipole component of the planetary magnetic field after several Gyrs. For Jupiter mass planets, \citet{reiners2010} predict that the magnetic field will be similar to the Jovian values (about 10G in the polar regions).

We note that \citet{reiners2010} used the planetary luminosity estimates from \citet{burrows1997} who assume a scenario where the planet gradually cools down as it ages. As we have discussed above, for most HJs such a cooling scenario is likely incorrect and the heat from the parent star plays a key role. Assuming that the interiors of inflated HJs are much more energetic than their hypothetical Jupiter-radius counterparts, the scaling law proposed by \citet{christensen2006} will predict larger magnetic field strengths. In this Letter we carry out the analysis by \cite{reiners2010} again to estimate the magnetic field strength in HJs. However, for planetary luminosity, we use the estimates by \citet{thorngren2017} who take into account the extra energy needed to inflate the planets. As we show below, after this correction, we can expect magnetic field strengths in HJs that may be an order of magnitude larger than the predictions by \citet{reiners2010}. Below we describe out methodology and present the results and the related discussion.

\section{Methods} \label{sec:method}
The studies by \citet{reiners2010} and \citet{thorngren2017} provide the basic concepts for our analysis. The former predicts the mean magnetic field in exoplanets using the luminosity and the latter provides estimates for the luminosity that are consistent with the radius inflation. Here, we only highlight the basic concepts and encourage the interested reader to refer to the original papers or more in-depth details.

\subsection{Mean magnetic field}
Using an ensemble of geodynamo simulations, \citet{christensen2006} inferred an empirical scaling law that relates the power generated by the buoyancy forces in the convection zone of planets with the mean magnetic field produced by the planetary dynamo. The scaling law was later tested and confirmed in dynamo simulations similar to giant planets with density stratified interiors \citep{yadav2013scaling1, yadav2013scaling2,schrinner2014}. In its original form, the scaling law requires the volume averaged buoyancy power, but \citet{christensen2009} showed that for a large class of objects (including giant planets and low mass stars) the buoyancy power can be related to the surface luminosity of the objects. Using these arguments, \citet{reiners2010} provide a simple scaling relationship to relate the mean magnetic field on the dynamo surface with the physical properties of plants/stars:
\begin{equation}
B^{dyn}_{mean} = 4.8\times10^{3}\,(M\,L^2)^{1/6}\,R^{-7/6}\,[G] \label{eg:Bdyn}
\end{equation}
where object mass $M$, luminosity $L$, and radius $R$, are normalized by the solar values. As \citet{reiners2010} note, in stars, $B_{dyn}$ represents the mean magnetic field on the surface, however, in giant planets, the dynamo surface is deeper in the planetary interior where conditions are favorable for metallic hydrogen formation. For higher mass giant planets, the dynamo surface comes closer to the surface. Due to the overlaying insulating layer in giant planets, magnetic field will be attenuated, especially the higher spherical harmonic degree modes, and we should account for this. \citet{reiners2010} assume that the depth of the dynamo surface is approximately inversely proportional to the planetary mass. They also assume that the mean dipole field strength is about half of the mean field strength on the dynamo surface. Under these assumptions, the value of the dipolar magnetic field on the planetary surface at  the rotational pole is 
\begin{equation}
B^{polar}_{dipole} = \frac{B^{dyn}_{mean}}{\sqrt{2}}\left(1-\frac{0.17}{M/M_J}\right)^3\,[G]. \label{eg:Bdip}
\end{equation}
This relationship assumes that the dynamo surface in Jupiter is at 0.83$R_J$. Note that the unquantifiable errors associated with the basic assumptions in Eqns.~\ref{eg:Bdyn} and \ref{eg:Bdip} are likely significant and, therefore, for simplicity, we do not track the errors in the field strength induced by the measurement errors in planetary mass and radius.

\subsection{Hot Jupiter luminosity}
For estimating the planetary luminosity, we refer to \cite{thorngren2017}. They compute the time-averaged incident flux on the planet from stellar and orbital parameters. Only planets with more than $2\times10^8 erg\,s^{-1}\,cm^{-2}$ incident flux, corresponding to the hot Jupiter population, are used. Furthermore, we restrict the data set to planets that are inflated as compared to Jupiter. For these planets, it should be safe to ignore the heat left over from formation as compared to the heat received from the star. Stars may also inject energy in close planets through tidal forces. However, the energy thus produced appears to be rather modest and short-lived in comparison to the persistent stellar irradiation \citep{leconte2010,thorngren2017}.

\cite{komacek2017b} discuss multiple heat-deposition scenario that are possible, and depending on the depth at which stellar heat is deposited, the amount needed to inflate HJs will vary. \cite{thorngren2017} assume that the heating uniformly affects the entire planetary adiabat, analogous to regime `2a' in \citet{komacek2017b}, in which heat is deposited at the center of the planet. To compute the anomalous heating \citet{thorngren2017} construct a functional relationship between the stellar heat flux and the heating required to reproduce the observed planetary radii. The Gaussian process model from that paper, combined with the incident flux on the sample planets, yielded the predicted heating.  We used the mean of the log efficiency prediction along with the nominal radius and flux to calculate the total heating in the planet. We then assume that this heating equals the planetary luminosity.

\begin{figure*}
\epsscale{1.16}\plotone{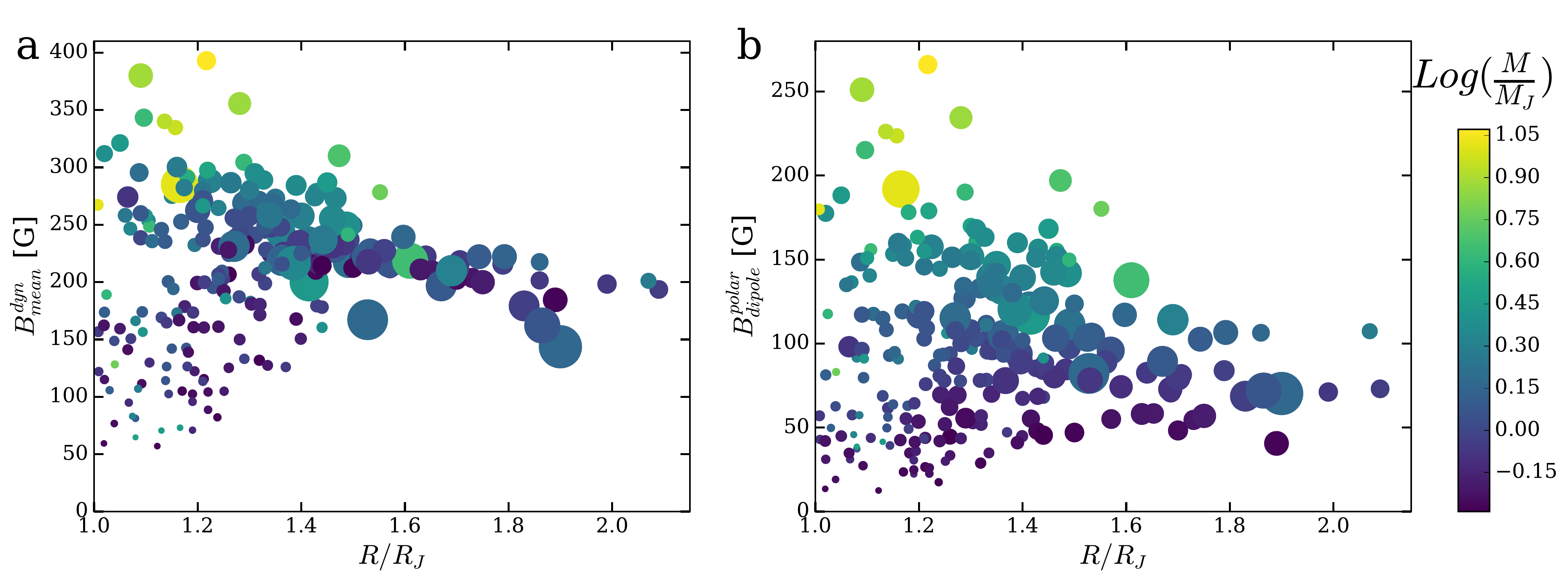}
\caption{Mean magnetic field strength on the dynamo surface ($B^{dyn}_{mean}$) versus the planetary radius in panel (a) and a similar figure for the magnetic field strength on the planetary surface at the rotational pole ($B^{polar}_{dipole}$) in panel (b). The symbol area scales as function of the incident stellar heat flux (in $erg\,s^{-1}\,cm^{-2}$) with the smallest and largest symbol representing $2\times10^{8}$ and $10^{10}$, respectively. The color of the symbols represent the mass of the planets. \label{fig:field}}
\end{figure*}

\section{Results}
Before we proceed, it is worth having the Jovian magnetic field in perspective. The morphology is largely that of a tilted dipole and its strength in the polar region is about 10G on the Jupiter's surface. Although, recent Juno measurements suggest that, on smaller length scales, Jovian surface field could be stronger and may reach values as high as 20G, and, at the deeper dynamo-surface, the smaller scale fields could be stronger than 50G \citep{moore2017}.

The magnetic field estimated from Eqns.~\ref{eg:Bdyn} and \ref{eg:Bdip} are plotted in Fig.~\ref{fig:field}. On the dynamo surface, the mean magnetic field is strongest in planets that are about 10-20\% inflated (Fig.~\ref{fig:field}a). Eqn.~\ref{eg:Bdyn} is roughly inversely proportional to the planetary radius, therefore, for a given planetary luminosity, if the radius is too large, then the larger surface area leads to weaker fields. Furthermore, for a given planetary radius, more massive planets will have stronger fields. Note the cluster of points in the lower left corner in Fig.~\ref{fig:field}a. For these planets with incident flux less than $10^9 erg\,s^{-1}\,cm^{-2}$, the radius inflation is modest due to low inflation efficiency \citep{thorngren2017}. Here, the primary trend is that of increasing field with increasing incident flux. For planets with higher incident fluxes, the effect of increased radius becomes important as well. The estimated mean magnetic field on the dynamo surface  reaches a maximum of about 400G in the most massive planet in our sample. Note that, as expected, a rough general trend where HJs receiving larger stellar heat exhibit larger mean magnetic fields is also evident from the figure. 

In Fig.~\ref{fig:field}b, the strength of the dipolar magnetic field on the rotational pole at the planetary surface is substantially attenuated as compared to the dynamo surface magnetic field in planets that have mass $\leq M_J$. For these planets, the overlaying insulating envelope filters out smaller scale magnetic fields and the dipole component also decreases. Nonetheless, since the field is rather strong due to larger luminosities of HJs, the surface fields in even planets with mass $<M_J$ generally exceeds that of Jupiter. For planets substantially more massive than Jupiter, the insulating envelope is smaller and the field is much stronger, and many such HJs are estimated to have about 150G polar dipole field. For the most massive of the HJs, the fields is estimated to be as high as 250G on the planetary surface. Note that for planets as massive as 10$M_J$, the small scale fields will be present close to the surface and we can probably expect field values reaching kilo gauss levels in localized regions. For the most inflated HJs, the maximum estimated field is in range 50 to 100G.

\section{Summary and Discussions}
In this study we show that in order to estimate the magnetic field strengths in hot Jupiters, it is crucial to take into account the energy that is responsible for inflating these planets. This energy is much larger as compared to the left-over energy from the planetary formation and it might be able to produce much stronger magnetic fields than the typical values in Jupiter. Our calculations suggest that in HJs with mass close to Jupiter we can expect magnetic fields that could be 10 times stronger than those in Jupiter. In more massive planets, the expected field strength increases to about 250G in HJs with  $\approx10M_J$ mass. 

We followed the study by \citet{reiners2010} and they do note that 100G level magnetic fields in Jupiter like exoplanets can be expected but only during very young ages when most of the heat from formation is still present. Since the radius inflation in HJs is a very good indicator that tells us that HJs have highly energetic interiors, the field strength we estimate in our study is largely a manifestation of the perpetual state of youth HJs are locked into due to the stellar heat.

If HJs can produce the kind of magnetic fields we estimate, then we may expect huge planetary magnetospheres that will extend the protective sphere around these planets where exomoons will be shielded from atmosphere eroding stellar wind. This may aid in sustaining the atmosphere around massive enough exomoons. Larger field strength will also lead to higher cutoff electron cyclotron frequencies. For example, assuming a relationship $f_{ce}$[MHz]=2.8$B^{polar}_{dipole}$[G] where $f_{ce}$ is cyclotron frequency \citep{farrell1999}, about 280MHz cutoff is expected for a 100G polar field.  Therefore, we encourage observers to explore radio emission of nearby hot Jupiter systems in higher frequency bands than traditionally used \citep{murphy2014, sirothia2014, griessmeier2015}. In this context, it is worth highlighting the recent analysis by \citet{weber2017} who suggest that based on earlier predictions for exoplanetary magnetic fields, the cyclotron-maser-instability radio emission will not be able to escape the magnetosphere of HJs with mass close to $M_J$. The five to ten fold higher magnetic field we predict from our analysis may be able to change the plasma conditions around HJs to such an extent that radio emission may escape. To quantify the changes, more analysis similar to \citet{weber2017} should be carried out for HJs with the higher magnetic fields we predict.

Given the semi-conducting nature of the atmosphere on the day-side of hot Jupiters, it is certain that the atmospheric circulation will be affected by the planetary magnetic field. Such an effect could be a simple drag force \citep{rauscher2013} or a more intricate one that may change the nature of the atmospheric circulation entirely \citep{batygin2013, rogers2014}. Studies so far have not ventured in to the field strength regime we propose here and it will be interesting to find out how atmospheric circulation changes under the strong field conditions.

We note that our predictions are solely based on the scaling law proposed by \citet{christensen2009} which does not assume any dependence on the rotation period of planets/stars as long as it is fast enough. This assumption largely aligns with the activity saturation phenomenon observed in cool stars \citep{reiners2009Saturation, wright2011, stelzer2016, astudillo2017, newton2017}. However, several studies also propose that even in the saturated activity regime, there might be some dependence on rotation rate with faster rotation leading to stronger field strengths/activity in stars \citep{jeffries2011, argiroffi2016, kao2016, shulyak2017}. If true and it extends to gas giants as well, then we may expect a trend in our estimates as a function of the rotation period. However, HJs are expected to be tidally locked with the parent star, and, therefore, the dependence on rotation might not be a large factor.

We also assumed that the heat is deposited in the planetary center, which may be a simplistic assumption. In a more realistic scenario, the heat is probably deposited off-center \citep{komacek2017b}. We expect that such off-center heating might disrupt coherent convective flows and/or magnetic field lines, perhaps leading to a less efficient planetary dynamo as compared to a center-heating scenario. Keeping this in mind, our estimated fields are then likely upper limits on the magnetic fields predicted by contemporary geodynamo simulations. To make these estimates more precise, we have to study the dynamo mechanism in hot Jupiter planets and model the effect of heat deposition at a range of off-center depths.

\acknowledgments

Authors thank Phil Arras and Thaddeus Komacek for interesting discussions, and the referee for a thoughtful review.


\begin{thebibliography}{}
\expandafter\ifx\csname natexlab\endcsname\relax\def\natexlab#1{#1}\fi

\bibitem[{Akeson {et~al.}(2013)Akeson, Chen, Ciardi, Crane, Good, Harbut,
  Jackson, Kane, Laity, Leifer, {et~al.}}]{akeson2013}
Akeson, R., Chen, X., Ciardi, D., {et~al.} 2013, Publications of the
  Astronomical Society of the Pacific, 125, 989

\bibitem[{Argiroffi {et~al.}(2016)Argiroffi, Caramazza, Micela, Sciortino,
  Moraux, Bouvier, \& Flaccomio}]{argiroffi2016}
Argiroffi, C., Caramazza, M., Micela, G., {et~al.} 2016, Astronomy \&
  Astrophysics, 589, A113

\bibitem[{Astudillo-Defru {et~al.}(2017)Astudillo-Defru, Delfosse, Bonfils,
  Forveille, Lovis, \& Rameau}]{astudillo2017}
Astudillo-Defru, N., Delfosse, X., Bonfils, X., {et~al.} 2017, Astronomy \&
  Astrophysics, 600, A13

\bibitem[{Batygin {et~al.}(2013)Batygin, Stanley, \& Stevenson}]{batygin2013}
Batygin, K., Stanley, S., \& Stevenson, D.~J. 2013, The Astrophysical Journal,
  776, 53

\bibitem[{Burrows {et~al.}(1997)Burrows, Marley, Hubbard, Lunine, Guillot,
  Saumon, Freedman, Sudarsky, \& Sharp}]{burrows1997}
Burrows, A., Marley, M., Hubbard, W., {et~al.} 1997, The Astrophysical Journal,
  491, 856

\bibitem[{Christensen(2010)}]{christensen2010}
Christensen, U. 2010, Space science reviews, 152, 565

\bibitem[{Christensen \& Aubert(2006)}]{christensen2006}
Christensen, U.~R., \& Aubert, J. 2006, Geophysical Journal International, 166,
  97

\bibitem[{Christensen {et~al.}(2009)Christensen, Holzwarth, \&
  Reiners}]{christensen2009}
Christensen, U.~R., Holzwarth, V., \& Reiners, A. 2009, Nature, 457, 167

\bibitem[{Farrell {et~al.}(1999)Farrell, Desch, \& Zarka}]{farrell1999}
Farrell, W., Desch, M., \& Zarka, P. 1999, Journal of Geophysical Research:
  Planets, 104, 14025

\bibitem[{Grie{\ss}meier(2015)}]{griessmeier2015}
Grie{\ss}meier, J.-M. 2015, in Characterizing stellar and exoplanetary
  environments (Springer), 213--237

\bibitem[{Grunblatt {et~al.}(2016)Grunblatt, Huber, Gaidos, Lopez, Fulton,
  Vanderburg, Barclay, Fortney, Howard, Isaacson, {et~al.}}]{grunblatt2016}
Grunblatt, S.~K., Huber, D., Gaidos, E.~J., {et~al.} 2016, The Astronomical
  Journal, 152, 185

\bibitem[{Hartman {et~al.}(2016)Hartman, Bakos, Bhatti, Penev, Bieryla, Latham,
  Kov{\'a}cs, Torres, Csubry, de~Val-Borro, {et~al.}}]{hartman2016}
Hartman, J.~D., Bakos, G.~{\'A}., Bhatti, W., {et~al.} 2016, The Astronomical
  Journal, 152, 182

\bibitem[{Jeffries {et~al.}(2011)Jeffries, Jackson, Briggs, Evans, \&
  Pye}]{jeffries2011}
Jeffries, R., Jackson, R., Briggs, K., Evans, P., \& Pye, J.~P. 2011, Monthly
  Notices of the Royal Astronomical Society, 411, 2099

\bibitem[{Kao {et~al.}(2016)Kao, Hallinan, Pineda, Escala, Burgasser, Bourke,
  \& Stevenson}]{kao2016}
Kao, M.~M., Hallinan, G., Pineda, J.~S., {et~al.} 2016, The Astrophysical
  Journal, 818, 24

\bibitem[{Komacek \& Youdin(2017)}]{komacek2017b}
Komacek, T.~D., \& Youdin, A.~N. 2017, arXiv preprint arXiv:1706.07605

\bibitem[{Laughlin {et~al.}(2011)Laughlin, Crismani, \& Adams}]{laughlin2011}
Laughlin, G., Crismani, M., \& Adams, F.~C. 2011, The Astrophysical Journal
  Letters, 729, L7

\bibitem[{{Leconte} {et~al.}(2010){Leconte}, {Chabrier}, {Baraffe}, \&
  {Levrard}}]{leconte2010}
{Leconte}, J., {Chabrier}, G., {Baraffe}, I., \& {Levrard}, B. 2010, Astronomy
  \& Astrophyscs, 516, A64

\bibitem[{Lopez \& Fortney(2016)}]{lopez2016}
Lopez, E.~D., \& Fortney, J.~J. 2016, The Astrophysical Journal, 818, 4

\bibitem[{Marley {et~al.}(2007)Marley, Fortney, Hubickyj, Bodenheimer, \&
  Lissauer}]{marley2007}
Marley, M.~S., Fortney, J.~J., Hubickyj, O., Bodenheimer, P., \& Lissauer,
  J.~J. 2007, The Astrophysical Journal, 655, 541

\bibitem[{Moore {et~al.}(2017)Moore, Bloxham, Connerney, J{\o}rgensen, \&
  Merayo}]{moore2017}
Moore, K.~M., Bloxham, J., Connerney, J.~E., J{\o}rgensen, J.~L., \& Merayo,
  J.~M. 2017, Geophysical Research Letters, 44, 4687

\bibitem[{Murphy {et~al.}(2014)Murphy, Bell, Kaplan, Gaensler, Offringa, Lenc,
  Hurley-Walker, Bernardi, Bowman, Briggs, {et~al.}}]{murphy2014}
Murphy, T., Bell, M.~E., Kaplan, D.~L., {et~al.} 2014, Monthly Notices of the
  Royal Astronomical Society, 446, 2560

\bibitem[{Newton {et~al.}(2017)Newton, Irwin, Charbonneau, Berlind, Calkins, \&
  Mink}]{newton2017}
Newton, E.~R., Irwin, J., Charbonneau, D., {et~al.} 2017, The Astrophysical
  Journal, 834, 85

\bibitem[{Rauscher \& Menou(2013)}]{rauscher2013}
Rauscher, E., \& Menou, K. 2013, The Astrophysical Journal, 764, 103

\bibitem[{Reiners {et~al.}(2009)Reiners, Basri, \&
  Browning}]{reiners2009Saturation}
Reiners, A., Basri, G., \& Browning, M. 2009, The Astrophysical Journal, 692,
  538

\bibitem[{Reiners \& Christensen(2010)}]{reiners2010}
Reiners, A., \& Christensen, U.~R. 2010, Astronomy \& Astrophysics, 522, A13

\bibitem[{Rogers \& Komacek(2014)}]{rogers2014}
Rogers, T.~M., \& Komacek, T.~D. 2014, The Astrophysical Journal, 794, 132

\bibitem[{Schneider {et~al.}(2011)Schneider, Dedieu, Le~Sidaner, Savalle, \&
  Zolotukhin}]{schneider2011}
Schneider, J., Dedieu, C., Le~Sidaner, P., Savalle, R., \& Zolotukhin, I. 2011,
  Astronomy \& Astrophysics, 532, A79

\bibitem[{Schrinner {et~al.}(2014)Schrinner, Petitdemange, Raynaud, \&
  Dormy}]{schrinner2014}
Schrinner, M., Petitdemange, L., Raynaud, R., \& Dormy, E. 2014, Astronomy \&
  Astrophysics, 564, A78

\bibitem[{Shulyak {et~al.}(2017)Shulyak, Reiners, Engeln, Malo, Yadav, Morin,
  \& Kochukhov}]{shulyak2017}
Shulyak, D., Reiners, A., Engeln, A., {et~al.} 2017, Nature Astronomy, 1, 0184

\bibitem[{Sirothia {et~al.}(2014)Sirothia, des Etangs, Kantharia, \&
  Ishwar-Chandra}]{sirothia2014}
Sirothia, S., des Etangs, A.~L., Kantharia, N., \& Ishwar-Chandra, C. 2014,
  Astronomy \& Astrophysics, 562, A108

\bibitem[{Spiegel \& Burrows(2013)}]{spiegel2013}
Spiegel, D.~S., \& Burrows, A. 2013, The Astrophysical Journal, 772, 76

\bibitem[{Stelzer {et~al.}(2016)Stelzer, Damasso, Scholz, \&
  Matt}]{stelzer2016}
Stelzer, B., Damasso, M., Scholz, A., \& Matt, S.~P. 2016, MNRAS, 463, 1844

\bibitem[{Thorngren \& Fortney(2017)}]{thorngren2017}
Thorngren, D.~P., \& Fortney, J.~J. 2017, arXiv:1709.04539

\bibitem[{Weber {et~al.}(2017)Weber, Lammer, Shaikhislamov,
  {et~al.}}]{weber2017}
Weber, C., Lammer, H., Shaikhislamov, I.~F., {et~al.} 2017, Monthly Notices of
  the Royal Astronomical Society, 469, 3505

\bibitem[{Wright {et~al.}(2011)Wright, Drake, Mamajek, \& Henry}]{wright2011}
Wright, N.~J., Drake, J.~J., Mamajek, E.~E., \& Henry, G.~W. 2011, The
  Astrophysical Journal, 743, 48

\bibitem[{Yadav {et~al.}(2013{\natexlab{a}})Yadav, Gastine, \&
  Christensen}]{yadav2013scaling1}
Yadav, R.~K., Gastine, T., \& Christensen, U.~R. 2013{\natexlab{a}}, Icarus,
  225, 185

\bibitem[{Yadav {et~al.}(2013{\natexlab{b}})Yadav, Gastine, Christensen, \&
  Duarte}]{yadav2013scaling2}
Yadav, R.~K., Gastine, T., Christensen, U.~R., \& Duarte, L. D.~V.
  2013{\natexlab{b}}, The Astrophysical Journal, 774, 6

\end{thebibliography}

\end{document}